¶ **These authors contributed equally to this work. All other authors are in alphabetical order.**

**Characteristics of ChatGPT users from Germany: implications for the digital divide from web tracking data**

**Celina Kacperski[1,2,¶] , Roberto Ulloa[1,3,*,¶], Denis Bonnay[4] , Juhi Kulshrestha[5] , Peter Selb[1], Andreas Spitz[6]**

[1] Cluster of Excellence "The Politics of Inequality", Konstanz University, Konstanz, Baden-Württemberg, Germany

[2] Wirtschaftspsychologie, Seeburg Castle University, Seeburg am Wallersee, Salzburg, Austria

[3] GESIS - Leibniz Institute for the Social Sciences, Cologne, Nordrhein-Westfalen, Germany[4] Department of Philosophy, Université Paris Nanterre, Paris, France

[5] Department of Computer Science, Aalto University, Espoo, Finland

[6] Department of Computer and Information Science, Konstanz University, Konstanz, Baden-Württemberg, Germany

[*] **Corresponding author**

**Email:** roberto.ulloa@uni-konstanz.de **(RU)**

[¶] **These authors contributed equally to this work. All other authors are in alphabetical order.**

# Abstract

A major challenge of our time is reducing disparities in access to and effective use of digital technologies, with recent discussions highlighting the role of AI in exacerbating the digital divide. We examine user characteristics that predict usage of the AI-powered conversational agent ChatGPT. We combine behavioral and survey data in a web tracked sample of N=1376 German citizens to investigate differences in ChatGPT activity (usage, visits, and adoption) during the first 11 months from the launch of the service (November 30, 2022). Guided by a model of technology acceptance (UTAUT-2), we examine the role of socio-demographics commonly associated with the digital divide in ChatGPT activity and explore further socio-political attributes identified via stability selection in Lasso regressions. We confirm that lower age and higher education affect ChatGPT usage, but do not find that gender or income do. We find full-time employment and more children to be barriers to ChatGPT activity. Using a variety of social media was positively associated with ChatGPT activity. In terms of political variables, political knowledge and political self-efficacy as well as some political behaviors such as voting, debating political issues online and offline and political action online were all associated with ChatGPT activity, with online political debating and political self-efficacy negatively so. Finally, need for cognition and communication skills such as writing, attending meetings, or giving presentations, were also associated with ChatGPT engagement, though chairing/organizing meetings was negatively associated. Our research informs efforts to address digital disparities and promote digital literacy among underserved populations by presenting implications, recommendations, and discussions on ethical and social issues of our findings.

# Introduction

A digital divide exists between individuals that have access to and can effectively use modern information and communication technologies and those who do not [1,2]. The concept of the digital divide emerged from empirical observations in the late 1990s about the growing inequality in access, use and knowledge of information technologies: it was initially centered around the disparities in internet access between different groups [3], and the correlations between these disparities and other inequalities [4], often including socio-demographic and structural characteristics, e.g., age, gender, income, education and residence type [5]. A large body of research on inequality related to internet usage postulates consequences such as missed economic and social opportunities, lack of added value from the online world, and an uneven distribution of benefits from digitalization [6]. Research on the digital divide has since expanded to include a myriad of digital tools [2,7], known as the second-level digital divide [8].

Artificial intelligence (AI) enters this conversation with the potential to both exacerbate and bridge the digital divide [9,10]. There have been calls to analyze sociotechnical systems critically, as those lacking access to AI technologies may miss out on the benefits and opportunities they offer, such as improved efficiency, automation, and supporting decision-making capabilities [11]. This might be particularly true for AI-powered conversational agents (CAs), whose range of applications continues to expand [12]. In the past year, CAs, exemplified by ChatGPT, have become increasingly capable of engaging in natural and dynamic conversations with users, offering a platform for users to interact, seek information, and even find companionship through virtual dialogue. While the use of ChatGPT and contemporary CAs has received much scientific attention [12], most of it has focused on the implications of CAs in very narrow thematic areas, e.g., questioning the consequences for healthcare [13] and medicine [14], education [15] and research [16,17], and the consequences to human decision-making and behavior [18–20].

So far, little is known about who the users of ChatGPT are, and how differences in access and usage are reflected across subgroups and socio-demographic strata. Recent reviews point to potential existing gaps in access, and suggest promoting digital literacy and empowerment to address hypothesized inequalities in access [21,22]. This is in line with similar arguments for strengthening AI literacy among underserved populations [23]. However, targeting underserved communities and groups first requires the identification of features associated with the usage of CAs. The present research contributes to this goal by uncovering disparities in using ChatGPT that are often related to the digital divide (age, education, full-time employment, and parenthood), while identifying novel dimensions (e.g., writing skills and political knowledge). We base our analysis on the combination of survey data with online behavioral data (i.e., web browsing traces), guided by a framework of technology acceptance (UTAUT-2). In light of our findings, we discuss implications, recommendations, and ethical and social issues for targeting the digital divide.

# Theoretical framework

The digital divide can be understood through the lens of theories of technology acceptance and usage [7]. Revising eight such theories, Venkatesh [24] formulated the Unified Theory of Acceptance and Use of Technology (UTAUT) and, later on, integrated additional constructs and relationships suitable for the consumer context, namely UTAUT-2 [25]. The latter comprises seven constructs that influence the acceptance of technology: (1) performance expectancy, i.e. a technology's usefulness; (2) effort expectancy, i.e. its ease of use; (3) hedonic motivation, i.e. its enjoyability; (4) facilitating conditions; (3) social influence; (6) price value; and (7) habit. These can be often moderated by age, gender and experience.

The UTAUT-2 has been deemed suitable to analyze the adoption and use of AI tools [26,27], and has already been applied in the study of the acceptance of conversational agents and chatbots in general [28,29], and in health care contexts in particular [30]. As a theoretical framework, it is well-suited to help clarify the relationship between the user characteristics that have been deemed important in the context of studying the digital divide and the use of AI-powered services such as ChatGPT. While our research focuses specifically on user characteristics, covering demographics, facilitating conditions and hedonic motivation, employing this theoretical frame also sets the foundation to better study acceptance of AI-powered services in future work.

Firstly, the UTAUT-2 explicitly defines in its base model age and gender and prior experience as variables that affect technology acceptance [25]. Ample research has shown that older people [31] and women [32] tend to have lower access to and acceptance of technologies [33]. Research also shows that individuals who do already have access to some technologies are more likely to continue using them and become more proficient [34,35], and that such experience is cumulative: accepting one technology positively affects performance and effort expectancies about novel technological innovations [36]. Education has been found to inconsistently moderate UTAUT-2 factors, though some evidence showcases its interaction with performance, effort expectancy, and facilitating conditions, with higher educational levels predicting higher intention to adopt novel autonomous technology [37,38] and innovative health solutions [39]. Higher educational attainment has been found to have lower effort expectancy regarding the internet [40]. Consequently, research has shown higher levels of education to be positively associated with digital literacy [41] and higher technology adoption [5].

Secondly, UTAUT-2 states that facilitating conditions, i.e., resources, infrastructure and support at disposal, are necessary for technology adoption [25]. Acceptance of AI and IoT services has been shown to be higher for individuals with higher socio-economic status [42,43], and having lower income has been shown to negatively affect performance expectancy [40]. Indeed, studies consistently find lower income as a negative predictor of internet access and online activity [44–46]. Although it is possible that the income gap is slowly decreasing [47], low-income households might have less experience with online technologies as a result of the delay. Rural areas are often considered a special case in research on technology acceptance [48,49]: urban and rural residence of individuals impact adoption of digital technologies in a complex, diverse and multidirectional way, mainly through infrastructure and public access provisions [50], acting as facilitating conditions: those living in rural areas are often less prone to using the internet [51]. Finally, time availability is a particular facilitating condition in technology acceptance models, so time constraints play a major role [52], for example when balancing full-time work [53] or care responsibilities [54,55] with the cognitive and time requirements for learning novel technologies. Hedonic motivation itself is also an important consideration: individuals with higher technological and general literacy and those that are intrinsically motivated have previously been shown to more readily adopt novel technologies due to enjoyment of these services [23,56,57].

In summary, in light of the UTAUT-2 framework and existing research, user characteristics including demographics and facilitating conditions that are prominent in the literature on the digital divide play a substantial role in the adoption and usage of digital technologies, and should be further investigated in the context of AI-powered tools, while pathways through which adoption is ensured need to be further explored [27].

# Research agenda

So far, if at all, the usage of AI-powered services has been studied using self-report measures, such as the AI literacy scale [23,57,58]. The limited instances of self-reported usage statistics of ChatGPT include one survey of U.S. adults, which reported 14% of respondents having used it [59], with a follow-up reporting 18%, indicating men and college-educated adults reporting more frequent usage [60]. And one survey of German citizens reported 27% of respondents were aware of ChatGPT, and 11% were using it, mostly in professional and educational contexts [61]. However, neither of these examples utilize behavioral data or investigate relevant user characteristics beyond base demographics.

Our research thus improves on prior work by merging (1) web tracking data, i.e., behavioral traces collected from participants while they are browsing the internet in a private context, and (2) survey data of their demographics and socio-political characteristics. This two-pronged approach allows us to contribute a study of the users of ChatGPT, i.e., to identify predictors of ChatGPT activity across a wide variety of user characteristics. We target three behavioral response variables by analyzing activity on ChatGPT (i.e., the domain chat.openai.com): usage, visit frequency, and adoption (weeks with visits to the domain, as a metric for long-term ChatGPT adoption). Based on prior literature on the digital divide [9,62], and within the UTAUT-2 framework [25], we pose the following hypotheses: (1) Being older will be associated with lower ChatGPT activity. (2) Being male will be associated with higher ChatGPT activity. (3) Higher educational attainments and more years of schooling will be associated with higher ChatGPT activity. (4) Higher income will be associated with higher ChatGPT activity. (5) Urban (compared to rural) residence will be associated with higher ChatGPT activity. (6) Experience with the internet, i.e. overall internet activity, will predict ChatGPT activity.

We also add the following exploratory analyses: we expect that any factors associated with time constraints (more work and care responsibilities) will likely reduce ChatGPT activity. We examine whether higher literacy (writing and media related) and usage of social media will be associated with higher ChatGPT activity. And finally, we investigate whether political knowledge, political participation both on and offline, and political leanings are related to ChatGPT activity.

# Methods

We used data collected for the Seek2Judge project [63], which was reviewed by the institutional ethics board of the University of Konstanz. Data was collected by European panel provider Bilendi GmbH, who obtained explicit informed consent from all participants before they signed up for the web tracking program and before the survey. We received pseudonymized web tracking data from December 1, 2022 until October 31, 2023, with trackers installed on personal computers and mobile devices. Specifically, for each participant, the data contains the visited URL, the time of visit and the pageview duration. The study participants were all above 18 years of age. We identified all visits to the web domain chat.openai.com, which hosted the conversational agent ChatGPT during the analyzed period.

From February 3 to 22 2023, we surveyed participants after gaining their written informed consent (via checkbox), and consent to match their survey responses to their web tracking data. The questionnaire included standard socio-demographic variables, as well as questions about social media usage and literacy, political participation, attitudes, and knowledge. For our analysis, we matched the browsing data of those participants with their survey responses and retained N = 1367 participants, of whom 11.78% (N=161) visited chat.openai.com. Due to a large number of relevant survey and observed variables (N=48), we mitigate associated issues such as multicollinearity by first conducting a feature reduction (via stability selection and Lasso [64]) and then fit regression models, including our hypothesized variables and those selected by the feature reduction. For collinearity checks, we include correlations (S1.2 Fig and S7.1.1 – S7.1.3) and variance inflation factor values (S7.2 Table). For completeness, we also provide results from zero-inflated models in the supplement (S8.1 and S8.2 Tables).

## Survey details

The survey was conceived in the context of the Seek2Judge project to measure a variety of socio-demographic and political variables in preparation for another, experimental study of political attitude formation as a result of online information search. As a result, the survey data does not contain any variables specific to LLM/AI or ChatGPT usage. Here, we report the survey constructs relevant to our final regression analyses. Due to space limitations, a full list of constructs and items as they were used in the survey and included in the LASSO feature selection (45 total) can be found in S1.1 Table, and a correlation table across all constructs (S1.2 Fig). We surveyed the following socio-demographics: age, gender, educational attainment, years of schooling, income, residence type (rural or (sub)urban), working situation (full-time or not), and number of children in the household. We used items from the resource model of political participation related to civic skills [65] to evaluate writing, attending meetings, organization and presentational skills, for example: “In which contexts have you carried out the following activity in the last 6 months: wrote a letter or email. Answer options: in the work context; in the private context; in the context of membership in political organizations; not performed at all in the above contexts.” [65] and 2 items from the Need for Cognition scale to evaluate need for cognition, for example “I like it when my life is full of tricky tasks that I have to solve.” [66]. Usage of social and new media was asked in a multiple-choice grid of commonly used platforms (e.g., Facebook, Twitter), and with the New Media Literacy scale, for example: “I know how to use search engines and research tools to get the information I need.” [67]. We also asked a variety of political interest variables, including knowledge of the political system (Example: “By whom is the Federal Chancellor of the Federal Republic (Germany) elected?”, single choice out of 4); and knowledge about politicians (Example: "Please assign Markus Söder to the correct party. Choices: CDU/CSU, FPD, SPD, Green Party, AfD, don't know.") [68], as well as political news use online ("How many days a week do you on average consume political news on the internet (all devices)?").

## Participants

We predefined the sample in the following manner: all panel participants signed up to the web tracking panel were eligible (N=7048, registered panelists with at least 1 visit in the previous 6 months); they had to agree to participate in the survey; and we excluded participants who failed our instructed response item attention check [69]. We finally report a sample of N=1367 German residents. Of these, N=512 self-identified as women (48.28%, 51.72% identified as men). The median age was 49 (M = 47.66, SD= 12.61, 18-80, with N=80 over the age of 65). Level of education was distributed as follows: N=176 (12.87%) finished elementary school, N=570 (41.70%) finished middle school, N=282 (20.62%) finished high school and N=339 (24.79%) held a university degree. Participants reported on average 14.25 years of schooling (median = 13, SD=3.5, min=2, max=38). The majority of participants N=1010 reported no children (median = 0), with the maximum number being 4 (for N=22) for a mean of M=0.38 (SD=0.72). N=288 (21.07%) of our participants were from Eastern Germany. In terms of their residence type, N= 190 were living in rural areas, N=535 lived in suburban areas and N=642 lived in urban areas. The median net income was 2000-3000 Euro, with N=128 households earning under 1000 Euro, and N=163 over 5000 Euro per month, while N=48 did not report their income, for which we control by adding an "income disclosed" variable. The majority of participants (N=737, 53.91%) were full-time employed. S2 Fig presents crossed age, gender, and education distributions.

## Web tracking dependent and independent variables

For our analyses, we used three response variables from the web browsing data to measure participants' activity on chat.openai.com: ChatGPT usage (dummy coded, 1 = visited chat.openai.com at least once); ChatGPT engagement: chat.openai.com visits (count, total number of visits to ChatGPT during the 11-month study period); and adoption (count, number of weeks that included visits to ChatGPT during the 11-month study period). The latter two variables distinguish between users that had short-lived but intensive interaction with ChatGPT (i.e., users who may have used it a lot in the span of a few weeks) and users that consistently interacted with it over the 11-month study period.

We also used three metrics of participants' internet activity as independent variables: (1) weekly web visits (logarithm of average number of weekly web visits, excluding those to chat.openai.com), (2) weekly web usage duration (logarithm of weekly average time spent on the internet), and (3) weekly activity, the number of weeks in which the tracking tool recorded at least one visit. Weekly web visits and weekly web usage duration are very highly correlated (r = 0.91), but we kept both for the LASSO procedure as they might have targeted different internet usage patterns (e.g., individuals that mostly watch videos online would display a high duration with low number of visits). We included weekly activity to control for inconsistent internet usage (due to, e.g., vacation periods) and potential measurement errors (e.g., technical issues with the tracking tool). Visits in all cases are aggregated visitation sequences of pages belonging to the same domain loaded within 30 minutes of each other (see S3.1 Text for details). These three metrics are included in the feature selection method together with the 45 survey variables.

## Analyses

The web tracking dataset was merged with the collected survey data. For analyses, for ChatGPT usage, all N=1361 participants were considered, and due to the binary nature of the data (coded 0,1 for non-usage and usage), we used logistic regression for binary classification. We report odds ratios as typically used in logistic regressions and when the outcome is rare (as in our case, only 11.77% of participants show usage), since it is a good estimate of the association between a variable and the probability of an event occurring [70]. For the two engagement variables, we excluded 0s and used the remaining N=161 participants who had at least 1 visit. As both are counts with heavy tails (see S3.2 Fig), we used the negative binomial and report incidence rate ratios, which measures the rate of cases occurring over time [71].

To better understand the effect of the main independent variables outlined in our hypothesis and to explore other potential explanatory factors, but reduce those factors to the most important ones, we conducted feature reduction using Lasso regressions [64] (R package glmnet [72]). Lasso is a suitable method to remove redundant variables to reduce model overfitting [73]. Nevertheless it suffers from instability in the selection of variables [74], e.g., it could replace variables that explain a similar amount of the data depending on the subsample used for the fit. To counteract this, we followed the stability selection method proposed by Meinshausen & Bühlmann [75,76]: for ChatGPT usage, we created 1000 samples without replacement of 50% (N=534) of our data; for each sample, we obtained selected features with 10-fold cross-validation, employing the best lambda for each sample. For visits and adoption, we created samples of 80% of the data (N=129) and we used 5-fold cross-validation to mitigate instabilities due to, otherwise, extremely small folds. For each dependent variable, the final set of features comprises those that were selected in at least 50% of the Lasso regressions. As Lasso disallows missing data, we imputed missing values (N=6 variables affected, 0.05-3% of values) using the R package missForest [77].

The described procedure was carried out for each of the three response variables. For better interpretability, we carried out regression analyses using the complete dataset on only these selected features, plus age, gender, education, income and residence type, for which we had defined hypotheses. Features were always standardized. We used R [78] for our analysis: glm (with binomial family) of the stats package for usage, glm.nb of the MASS package for visits and adoption [79]. For the LASSO stability selection method of visits and adoption, we used cv.glmregNB of the mpath package [80,81].

As a measure of predictive power for ChatGPT usage, we report the mean of the AUCs (Area Under the Receiver Operating Characteristic (ROC) curves [82] of the 1000 repetitions, for which we employ the left-out 50% as a test dataset in each repetition. For regressions predicting the three response variables of ChatGPT activity, we report regression model coefficients (odds and incidence rate ratios) and significance at the 5% level.

# Results

Participants had a median of 3754 web visits across 11 months (M=6905.3, SD= 9197.87), with the most active participant logging 89712 visits, while the least active logged 1. N=161 of our participants visited chat.openai.com, which constitutes 11.77% of our sample (N=1206 did not). For those that did visit ChatGPT, the median number of visits was 4 (M=20.58, SD=50.03, min=1, max= 400), and the number of weeks throughout which ChatGPT was visited was 3 (M=5.67, SD=7.41, min=1, max=42); see S3.2 Fig for distributions. Fig 1 presents domain visits to ChatGPT as a percentage of total web visits for all participants, across the time from December 1 2022, the day after ChatGPT launched, until the end of October 2023, our final data collection point.

**Fig 1. Percentage of ChatGPT visits.** X-axis shows bi-weekly data batches. Y-axis is the percentage of ChatGPT visits of total web visits.

## Feature selection

Fig 2 summarizes the results of the feature selection procedure (left plot). Of the 48 features, 27 features were selected, with 3 of them being across all three response variables. S5.1.1 – S5.3.3 Fig holds histograms of iterations in which features were selected, examples of cross-validation and coefficient plots. For ChatGPT usage, the mean AUC of the 1000 repetitions is .78 (SD=.02). Dropping the three variables of internet activity as obvious strong predictors, the mean AUC is .66 (SD=.02) calculated over a 1000 iterations of stability selection. Example ROC plots can be found in S6.1 Fig and S6.2 Fig.

**Fig 2. Summary of the feature selection procedure.** The gradient of plot A (left) represents the number of repetitions (out of 1000) in which each feature was selected in the cross-validation Lasso regression (10-fold for usage; 5-fold for visits and adoption). Plot B only displays those that were selected in more than 500 repetitions, i.e., the features selected for regression models. Direction of association is colored red (negative) or blue (positive).

## ChatGPT usage and engagement (visits and adoption)

Fig 3 displays regression coefficients with 95% confidence intervals. The left plot displays odds ratios for ChatGPT usage, the middle and right plots display incidence rate ratios for variables associated with ChatGPT visits and adoption, i.e. the number of weeks with ChatGPT visits. Full regressions for each predictor are found in Table 1. The regressions include all features with one exception, namely web engagement for ChatGPT visits, because it is highly correlated with web activity (See S7.1.1 – S.7.1.3 Fig) and when included together in the regression, they display VIF values (4.11 and 4.23, see S7.2 Table) close to the threshold (5) that indicates potential multicollinearity. Results obtained for non-imputed data (see S4.1 Table and S4.2 Fig) and zero-inflated models (S8.1 and S8.2 Tables) were highly similar.

**Fig 3. Odds ratio and standard estimates.** The odds ratio and incidence rate ratio are re-calculated in the complete dataset for better interpretability. The three plots show odds ratio of the regression for ChatGPT usage (left), the incidence rate ratio for the visits (middle) and the incidence rate ration for adoption (right), all at 95% confidence intervals. The labels on the Y-axis are sorted decreasingly according to the odds ratio or the incidence rate ratio (X-axis).

**Table 1. Regression models for usage, visits and adoption of ChatGPT.**

| | Usage | | | Visits | | | Adoption | | |
|---|---|---|---|---|---|---|---|---|---|
| *Predictors* | *Odds Ratios* | *CI* | *p* | *Incidence Rate Ratios* | *CI* | *p* | *Incidence Rate Ratios* | *CI* | *p* |
| (Intercept) | 0.05 | 0.04-0.07 | **<0.001** | 5.72 | 4.21-7.91 | **<0.001** | 2.55 | 1.94-3.36 | **<0.001** |
| gender | 1.17 | 0.95-1.44 | 0.144 | 1.18 | 0.92-1.51 | 0.122 | 1.06 | 0.91-1.23 | 0.468 |
| age | 0.51 | 0.41-0.63 | **<0.001** | 0.73 | 0.59-0.89 | **0.002** | 0.77 | 0.67-0.88 | **<0.001** |
| education | 1.23 | 1.02-1.48 | **0.028** | 0.89 | 0.73-1.10 | 0.192 | 0.98 | 0.86-1.12 | 0.732 |
| income | 1.09 | 0.89-1.34 | 0.384 | 0.95 | 0.77-1.19 | 0.643 | 0.95 | 0.82-1.10 | 0.476 |
| residence rural | 1.19 | 0.99-1.42 | 0.054 | 1.04 | 0.86-1.27 | 0.713 | 1.07 | 0.93-1.22 | 0.339 |
| income disclosure | 0.79 | 0.67-0.94 | **0.007** | 0.82 | 0.68-0.97 | **0.011** | 0.85 | 0.75-0.95 | **0.002** |
| children in household | 0.81 | 0.65-0.99 | **0.043** | | | | | | |
| employment | 0.80 | 0.65-0.99 | **0.042** | | | | | | |
| new media literacy | 1.24 | 0.99-1.56 | 0.069 | | | | | | |
| SM Telegram | 1.14 | 0.94-1.38 | 0.186 | | | | | | |
| SM Facebook | 0.80 | 0.65-0.98 | **0.031** | | | | | | |
| SM WhatsApp | | | | 1.18 | 0.92-1.48 | 0.165 | | | |
| SM variety | 1.38 | 1.10-1.73 | **0.006** | 1.32 | 1.09-1.61 | **0.004** | 1.18 | 1.03-1.35 | **0.023** |
| political knowledge | 1.41 | 1.13-1.77 | **0.002** | | | | | | |
| political leaning | 0.84 | 0.69-1.01 | 0.065 | | | | | | |
| voted in election | 1.32 | 1.03-1.77 | **0.042** | | | | | | |
| political debate online | 0.82 | 0.67-0.99 | **0.047** | | | | | | |
| political debate offline | | | | | | | 0.88 | 0.76-1.03 | 0.095 |
| political online action | | | | 1.26 | 1.03-1.55 | **0.019** | 1.25 | 1.08-1.47 | **0.002** |
| political self-efficacy | | | | 0.65 | 0.47-0.90 | **0.001** | | | |
| religious membership | | | | 0.89 | 0.73-1.09 | 0.190 | 0.84 | 0.73-0.97 | **0.018** |
| need for cognition | | | | 1.40 | 1.10-1.79 | **0.002** | 1.18 | 1.00-1.40 | **0.045** |
| writing skills | | | | 1.47 | 1.14-1.89 | **0.001** | 1.22 | 1.02-1.47 | **0.025** |
| presentation skills | | | | 1.15 | 0.93-1.44 | 0.161 | | | |
| organization skills | | | | | | | 0.73 | 0.60-0.88 | **<0.001** |
| meeting skills | | | | | | | 1.23 | 1.03-1.47 | **0.020** |
| web duration | 2.53 | 1.90-3.41 | **<0.001** | | | | | | |
| Weekly activity | 2.41 | 1.72-3.53 | **<0.001** | | | | 1.31 | 0.91-1.88 | 0.116 |
| web visits | | | | 1.98 | 1.41-2.78 | **<0.001** | 1.36 | 1.07-1.74 | **0.012** |
| Observations | 1367 | | | 161 | | | 161 | | |
| $R^2$ Tjur | 0.214 | | | 0.630 | | | 0.573 | | |
| AIC | 789.480 | | | 1164.763 | | | 866.428 | | |
| AICc | 790.044 | | | 1169.043 | | | 871.245 | | |

**Note**: Predictors are sorted by thematic area, with the first six predictors included in all models due to pre-existing hypotheses. Empty rows indicate the feature was not selected by LASSO and is therefore not included in the regression models. P-values below the significance threshold are bolded.

First, we examined socio-demographic predictors common in the digital divide literature: in line with our hypothesis, being older was associated with lower ChatGPT activity across all three response variables. Higher levels of education were also positively related with ChatGPT usage, although we did not find a significant relationship with engagement. Surprisingly, we could not confirm expected effects for gender and income. We also did not find a significant relationship with residence type, however, contrary to our hypothesis, the direction of the effect indicates that residing in a rural area (as compared to (sub)urban living) might be associated with stronger ChatGPT engagement.

Following this, we examined the remaining variables in an explorative manner. Here, in line with our expectations regarding variables associated with time constraints, full-time employment and having more children were found to be significant negative predictors of ChatGPT activity. We also examined variables related to social media, where results were mixed: usage of a variety of social media sites was positively associated, while Facebook usage was negatively associated with ChatGPT activity. New media literacy, while Lasso selected as important for usage, showed no significant relationship, though the effect direction indicates the expected positive association.

In terms of political variables, we found political knowledge and voting in the election to be positive predictors of ChatGPT usage, and engaging in political online action of more ChatGPT engagement. However, engaging in online political debates as well as political self-efficacy was negatively associated, the first with usage, and the second with ChatGPT adoption over time. We did not find a significant relationship for political leaning or for engaging in offline political debates. Being a member of a religious organization was associated with lower ChatGPT adoption over time, but not with visits, and had not been selected for usage.

We also found some variables to be specifically and only associated with ChatGPT engagement. We found need for cognition to be positively associated with both visits and adoption. The same was true for individuals that reported using their writing skills more often. The effect of attending meeting and organizing meeting skills was only significant for visits, but not for longer-term adoption, with attending meetings positively associated, but organizing/chairing meetings negatively associated with visits.

Finally, as expected, internet activity metrics were the strongest predictors for all three: ChatGPT usage, visits and adoption over time. The number of active weeks as well as duration spent online predicted ChatGPT usage; and the number of overall internet visits predicted visits and adoption.

# Discussion

We examined variables associated with ChatGPT activity, leveraging a combination of web tracking and survey data. We examined basic socio-demographic variables strongly relevant in the literature on the digital divide [5,6,62] and relevant in the unified theory of acceptance and use of technology [25], as well as those selected via Lasso feature reduction. We found significant relationships for two of the former: age was a significant negative predictor of ChatGPT activity, indicating that younger individuals are more likely to start using it, and use it more when they do, both in terms of visits, and adoption over longer periods of time; and years of education was a positive predictor of ChatGPT usage, but not engagement, indicating that once individuals used it at all, they did not differ on the number of visits or length of adoption. We could not confirm that being male or having higher income would be associated with higher ChatGPT activity. This might be attributable to the inclusion of more nuanced factors that might better account for variance, such as writing and political knowledge; it remains to be seen whether solely relying on demographics as predictors may overshadow the growing relevance of other determinants of today's technologically multifaceted landscape.

Contrary to the hypothesized higher ChatGPT activity for urban residents that has been shown for internet activity previously [51], we did not find any significant effect of residence, though if anything, the effect seems to be negative, i.e. with rural inhabitants more likely to initiate usage. This is surprising, as even post-COVID, there has been no evidence that internet usage itself has shifted between rural and urban residents in Europe [83]. In our sample, we did not find a significant correlation between residence type and internet activity overall, so in the absence of this relation, our original assumption of residence type as a facilitating condition might not apply. If future research should find higher usage in rural regions, this might be country specific (Germany in our study), as the digital divide has been specifically targeted here to increase rural access to technologies [83]. Another possible explanation for such an effect could be that rurally residing citizens' reduced access to educational and entertainment resources means that, for their leisure usage, ChatGPT is more attractive and relevant, i.e. mapping onto the hedonic motivation pathway in the UTAUT-2.

The intuitively expected negative relationship between ChatGPT activity and underaged children in the household finds support in our data, even when accounting for total internet activity. On top of the time constraints, which we had mapped onto the facilitating condition pathway in the UTAUT-2, households with children might have concerns regarding ChatGPT's content relevance, unreliability, and privacy issues, which would be mapped onto or be precursors to performance expectancy. For full-time employees, better facilitating conditions might be available at work, and performance expectancy might be higher in that context: respondents could be using ChatGPT for work-related reasons, and avoid additional interactions for personal usage.

The connection between social media and ChatGPT usage appears to be complex. Our results indicate using a variety of social media channels is positively associated with activity, likely due to higher exposure to quantity and diversity of online environments – the social influence pathway of the UTAUT-2 might be at play here, with social contagion responsible for starting usage and continuous engagement. However, Facebook seems to be a distinct environment, where higher usage is related to less ChatGPT activity. It will be fruitful to investigate social media usage further to untangle these relationships.

We show a relationship between ChatGPT activity and various political activities as well as knowledge; political context hasn't received much attention in the conversation around digital divide [84]. Given that we control for socio-demographics often associated with political knowledge (e.g., education, age, gender), hedonic motivation might be the pathway that explains ChatGPT adoption for knowledge, i.e., political knowledge as a proxy for an underlying interest in information discovery and the development of rhetoric related to the current regulation debates. However, results for other political variables are not consistent: we find voting in elections and political online action to be positively associated with usage with ChatGPT and engagement respectively, but engaging in political debate (especially online) and political self-efficacy negatively predict ChatGPT usage and adoption respectively. Future studies should investigate this pattern more thoroughly: collecting chat history with CAs could help uncover more about how individuals search out political dialogue in this context. While researchers have started to look into ChatGPT's political bias [85,86] and its effects on political attitude formation [87] we could not identify any research on pre-existing differences in political self-perception and behavior and how this might affect CA usage, especially conversational contents. This might have implications for political opinion formation and political behaviors down the line.

Finally, literacy-related and organizational activities and skills, including need for cognition and items from the resource model of political participation [65], should be a strong focus in future evaluations: they seem to play a role for visits and duration spent, but not for usage as a metric. This is interesting as it suggests that while the initial level of some skills (such as writing) might not matter, continuous ChatGPT use requires a certain skill level – this would be in line with both the facilitating condition and habit pathways in the UTAUT: a certain skill level is necessary to experience major benefits, and habitual usage of the internet has been shown to increase internet usage in the past [34,35]. At the same time, the reverse could apply: ChatGPT is more relevant for people who require writing as part of their everyday lives, including note-taking or summarizing meetings: this would potentially increase their performance expectancy towards LLMs. In this regard, the relationship between personal innovativeness and behavioral intention to use AI systems has previously been established and might also explain the increased usage [88]. We also find a negative association between organization skills and adoption – the measure specifically targets the frequency with which individuals chair and organize meetings, so this relationship might be because, such skills or the roles of participants who carry them out, do not directly benefit from the technological capabilities of ChatGPT; this might be related to the time constraints evaluated in combination with work and care responsibilities, and showcase a facilitating condition in the context of the UTAUT-2.

## Limitations

The following limitations should be noted: firstly, our data is correlational, so causal relationships cannot be established, and it is possible that other factors and interactions are at play that could serve as alternative explanations of the relationships that we observe. As the survey was carried out in the beginning of February, it is, however, unlikely that reverse relationships could be responsible and ChatGPT might have affected more malleable variables in our dataset such as political knowledge or leanings.

Secondly, our data showcases ChatGPT usage as a relatively rare event, as only ~12% of our participants were found to have initiated an activity with ChatGPT, which means our sample size of people with data on ChatGPT usage is relatively small at N=161. It should be kept in mind that collecting web traces is a notoriously difficult undertaking, with many caveats and limitations previously discussed in the literature (for an overview, see [89]), including the fact that browsing data largely remains in the hands of private industry actors, and that research on such topics as novel technology adoption is additionally impeded by the impossibility of predicting when a disruptive technology will emerge. That said, while absence and presence of usage in this case are equally important to establish relationships with associated variables, the distribution of the two should be considered when interpreting and generalizing from our data. The rarity of the events particularly affects the sample size of the visits and adoption analysis; although LASSO is designed to select variables even in circumstances in which they exceed the number of observations (i.e., the "p > n" problem) and handling multicollinearity, we had to relax the conditions of the stability selection and cross-validations to choose the variables. Nevertheless, the selected variables were neither correlated nor did they display variable inflation factor (VIF) values above the recommended thresholds (S7.2 Table), nor do we make any claims regarding the discarded variables. While we are confident that our analysis well describes the patterns in our data captured by state-of-the-art behavioral measures of internet browsing, future web tracking studies should most definitely attempt to gather larger numbers of participants.

Third, our sample self-selected into the web tracking panel, so it is possible that it constitutes a particular subset of the population; results should be generalized with caution. It is unlikely that this would affect social desirability responses that would directly interact with ChatGPT activity, as participants were never primed about this focus of investigation. Sample demographics should be kept in mind: our participants were German, and though well-distributed across age, gender and education, this affects the generalizability of our findings. However, scholars from other countries may still find the framework we provide valuable to identify digital divide aspects in the context of the UTAUT-2 and learn from our methodological approach to gain a nuanced understanding of AI service usage patterns. Our identification of socio-political attributes and behaviors as predictors of usage extends beyond traditional demographic factors and could also be adopted, replicated, and expanded by other scholars. There might also be an inherent representativeness issue in studies that use volunteers from access panels in online behavior - if those group characteristics affect both panel membership and the behavior of interest, studying panelists might induce something akin to overcontrol bias [90], controlling away parts of the effect which group characteristics have on online behavior that runs through panel participation.

Fourth, it should be noted that trackers were installed on personal computers and mobile devices - while individuals might use their personal devices for office work as well, our findings speak mostly for usage during leisure and personal time - usage of ChatGPT in professional contexts will require further investigation.

Finally, as the survey was launched in the context of another study, we did not have the opportunity to include items in the survey regarding ChatGPT usage, the usage of other CAs, or propensity towards AI in general, to correspond our behavioral metric with self-reported measures. It is of course possible that the number of users of AI-powered conversational agents is higher or lower in non-panel populations, or even that participants used other CAs; however, considering the media prominence and market dominance of OpenAI even towards the end of 2023 with over 60% (competitors Character.AI, Bard and Perplexity collectively hold 19%) [91], it is likely that the number of participants who used a different CA and never opened ChatGPT is quite low. This is particularly because ChatGPT's main competition, Google Bard only became available in July in the EU and Google Gemini did not become available in Germany until December 2023. Future studies that leverage web browsing data to study the adoption of novel technologies such as ChatGPT could showcase these patterns by including such items in their surveys.

## Social and ethical issues and implications

While conversational AI services, such as ChatGPT, are revolutionizing the way individuals interact with the internet, a multitude of social and ethical challenges remain and require in-depth attention [21,22,92]. In terms of social challenges, addressing digital inequalities necessitates a focus on the affordability and accessibility of AI technologies like ChatGPT, ensuring they are economically feasible and technologically available across diverse socio-economic strata. AI services are already transforming job markets, from job losses in the technology sector to changing skill requirements that includes proficiencies in AI tools, to the emergence of new types of job [93]. Our behavioral data, in line with previous self-report surveys [59–61] suggest that only a small percentage of internet users have used ChatGPT at least once. Thus, there is a necessity for underserved populations to quickly acquire proficiencies with these technologies, lest they risk being left behind. At the same time, it is also important to consider regional differences to account for specific technological and cultural contexts as well as the peculiarities of pre-existing digital divides [48]. For example, similar usage of ChatGPT in rural as well as urban areas could be in part due to a highly developed ICT infrastructure in Germany [91% access, see 94], while it dispels the prejudice of rural residents as inherently less willing to accept technologies [95]. Other regions might fare differently.

To effectively bridge this divide, it is essential to implement targeted interventions such as digital literacy programs tailored for various age groups and digital proficiency levels, especially focusing on older and less digitally savvy individuals. Research should focus on why and how access is limited here, and best practices for usage. The initiatives should be accompanied by the creation of safe spaces, for example in libraries, in which users can freely access and experiment with AI services and tools while having the support of educational resources. Such spaces should optimally be sponsored by AI companies who can benefit from the promotion of their product, but more importantly, from a diverse set of experiences emerging from vulnerable populations.

Ethical aspects of AI have centered around transparency and accountability, privacy, anti-discrimination and justice, and safety [96], for many of which, arguably, technical solutions are partially available [97]. Implications for the digital divide are manifold: biases in AI algorithms, like those affecting women or minorities, are well-documented (for a review, see [98]), albeit still poorly quantified for large language models [99], stem from the data used to train algorithms, and can exacerbate inequalities [100]. The AI field also lacks diversity: Hagendorff highlights this within the AI community, pointing out that this leads to spaces and algorithms that do not represent a diverse user base [97]. Uninvolved communities might find the resulting tools technologies inapplicable or counterproductive to their contexts and needs. Regulations to improve team diversity and mandate ethical training for AI professionals, not unlike mandatory ethical courses in fields such as medicine, could help improve on these shortcomings.

The trust users place in information from online sources, starting from search engines [101] but now proliferated to conversational agents [102] is well-documented. Tools like ChatGPT are known for propagating biased and hallucinated information [15], which poses a particular challenge: according to our results, current ChatGPT users more likely include individuals with higher education, political knowledge, and engagement in writing activities. This demographic might trust

ChatGPT for information due to their intellectual curiosity and reliance on digital tools for work and communication. Despite their education and awareness, they may not fully grasp the limitations and biases of AI technologies. Trust in AI for analytical thinking and information gathering is concerning if it decreases the usage of diverse, critical sources and fosters reliance on potentially flawed AI-generated content. As we also find users of ChatGPT to be more active on social media, the spread of such generated content online might be accelerated. Educational resources, awareness campaigns, and accessible tools that provide sources with each claim might increase the likelihood for better understanding and responsible usage for the entire user base.

Finally, as AI companies collect copious amounts of user data, including users' interactions during the chats to further improve their services, privacy concerns abound [15,21]. Our research shows that those with more internet experience are the early adopters; however, some evidence shows that they are not necessarily well-informed about the dangers of sharing information with ChatGPT [29,103]. Tightening regulations to prevent the use and misuse of sensitive data while ensuring continuous feedback to users will therefore be paramount: privacy by default should be the norm, neither optional nor conditional on basic features of the service as is currently the case with for example ChatGPT history access.

In summary, the characteristics that play a role for ChatGPT activity should be evaluated taking into consideration both barriers and motivators to usage, and further investigation mapping them onto the UTAUT-2 framework might yield important insights to tackle the digital divide. Some, such as older individuals, those with less education and political knowledge, those who already use the internet less, those with less time due to full-time work and children, and those with fewer writing and social media skills, may be limited in their ability to access information, resources, and opportunities that emerge from usage of conversational agent technologies. As these technologies develop and become more important as everyday skills, the barriers in developing the necessary skills for good usage will only increase, and the possibility has been discussed that a part of the population will lag behind in a variety of educational and employment opportunities [21,23]. On the other hand, it is important to note that some of the characteristics we examine, such as residents of rural regions, individuals with more political knowledge and those with more need of writing, might reflect not just an absence of barriers, but also the existence of needs and affinities in the individuals that possess them. Future research should distinguish between these two approaches to tease apart where the need for intervention is most dire. Efforts should focus on bridging gaps by promoting the adoption of CAs to all individuals, and more research should focus on why and how access is limited, including theory building on how these factors play together. Additionally, best-practices for usage need to be better established, which includes research to evaluate the impacts and consequences of CA usage, not just in thematically restricted areas, but also for the individuals using them [15,19].

## Conclusion

Our research finds that ChatGPT users are more likely to be younger, better educated, with higher needs for cognition, make more frequent use of communication/civic skills (such as writing and holding presentations) and higher political knowledge. Other political variables, such as those measuring political participation (voting, debates) and self-efficacy, seem to have a more complex interaction with ChatGPT adoption. ChatGPT users are also more likely to be active on the internet and engage with a diverse array of social media platforms and less likely to have children and full-time jobs. Considering that conversational agents like ChatGPT can either exacerbate or mitigate the digital divide, these insights are crucial to better address the efforts in bridging this gap. Our results suggest that these efforts require inclusive and accessible technologies, and a multi-faceted approach that considers infrastructure, financial and time affordability, digital literacy, and political, cultural, social and ethical considerations. Furthermore, providing training and educational opportunities is essential to empower individuals and communities to effectively utilize and derive benefits from such services.

# Acknowledgments

The authors thank colleagues at Bilendi for providing access to the data. We thank the reviewers for their insightful comments and helpful feedback.

# Supporting information

Paper_chatgptusers_Supplementary Materials_final.pdf

Web tracking details.pdf